# Neutral-current Hall effects in disordered graphene

Yilin Wang,[1,2,4] Xinghan Cai,[2] Janice Reutt-Robey[1,4] and Michael S. Fuhrer[1,2,3*]

[1]*Materials Research Science and Engineering Center, Department of Physics, University of Maryland, College Park, MD 20742, USA*

[2]*Center for Nanophysics and Advanced Materials, University of Maryland, College Park, MD 20742, USA*

[3]*School of Physics, Monash University, Victoria 3800, Australia*

[4]*Department of Chemistry and Biochemistry, University of Maryland, College Park, MD 20742, USA*

* michael.fuhrer@monash.edu

A non-local Hall bar geometry is used to detect neutral-current Hall effects in graphene on silicon dioxide. Disorder is tuned by the addition of Au or Ir adatoms in ultra-high vacuum. A reproducible neutral-current Hall effect is found in both as-fabricated and adatom-decorated graphene. The Hall angle exhibits a complex but reproducible dependence on gate voltage and disorder, and notably breaks electron-hole symmetry. An exponential dependence on length between Hall and inverse-Hall probes indicates a neutral current relaxation length of approximately 300 nm. The short relaxation length and lack of precession in parallel magnetic field suggest that the neutral currents are valley currents. The near lack of temperature dependence from 7-300 K is unprecedented and promising for using controlled disorder for room temperature neutral-current electronics.

Graphene with spin degeneracy $g_s = 2$ and valley degeneracy $g_v = 2$ allows the possibility of charge-neutral currents of spin or valley polarization, and such currents have been the subject of intense theoretical and experimental interest [1-11]. Their experimental manifestation is a neutral (spin, valley) Hall effect in which a charge current $j$ drives a transverse neutral current $j_n$ of either spin ($j_s$) or valley ($j_v$) polarization. Graphene is not expected to spontaneously break inversion or time-reversal symmetries, and has very weak spin-orbit coupling [12-14]. Hence the interaction between charge currents and spin and valley currents in perfect graphene is negligible. However, various schemes have been used to realize strong coupling between charge and neutral currents in graphene. Neutral currents in graphene with $sp^3$ chemical bonding [6, 7] were interpreted as a spin Hall effect (SHE) due to enhanced spin-orbit coupling, though a recent report rejects that interpretation [11]. Inversion symmetry breaking through interaction with a substrate [8] or via perpendicular electric field in bilayer graphene [9, 10] can induce a valley Hall effect (VHE). However as-fabricated graphene devices have not been expected to exhibit neutral Hall effects, as they lack strong spin-orbit coupling, or any global inversion symmetry breaking.

Here we search for neutral Hall effects in graphene on silicon dioxide as-fabricated and after deposition of $5d$ electron-containing Au and Ir. Ir adatoms have been predicted to add strong spin-orbit coupling in graphene [15], and Au adatoms have been experimentally implicated in neutral Hall currents attributed to spin-orbit coupling [16]. We use the non-local Hall bar geometry [3, 5, 7, 8, 16] in which a non-local resistance signal $R_{NL}$ is generated through neutral Hall and neutral inverse Hall effects. Surprisingly, we observe a clear and reproducible neutral Hall effect in as-fabricated graphene. On deposition of Au and Ir we observe a reduction of the charge carrier mobility consistent with a small charge transfer from Au/Ir to graphene. The neutral Hall angle $\gamma_n = j_n/j$ exhibits a complex but reproducible dependence on gate voltage and disorder, and notably breaks electron-hole symmetry. An exponential dependence on length between Hall and inverse-Hall probes indicates a neutral current relaxation length of approximately 300 nm. We observe no sign of a spin-orbit gap opening in graphene down to a temperature of 7 K, and a negligible dependence of $R_{NL}$ on parallel magnetic field, inconsistent with the precession expected for spin currents. The observations rule out spin as the neutral current type. Instead we conclude that that the neutral Hall effect in both as-fabricated and adatom-decorated graphene is a disorder-induced VHE. The near lack of temperature dependence from 7-300 K is unprecedented for a VHE in graphene [8-10] and promising for using controlled disorder for room temperature valleytronic devices.

Figure 1(a) shows a typical atomic force microscopy (AFM) image of our Hall bar geometry. A charge current, $I_{NL}$, injected across one cross of the Hall bar, generates a nonlocal voltage drop, $V_{NL}$, transverse to another cross of the Hall bar. The nonlocal resistance is $R_{NL} = V_{NL}/I_{NL}$. The (local) longitudinal resistivity $\rho_{xx}$ is measured in the usual manner. Graphene flakes are obtained by mechanical exfoliation of graphite on a 300-nm-SiO$_2$/Si substrate and the devices are fabricated in two electron beam lithography steps to establish Cr/Au (5 nm/100 nm) electrodes and define the Hall bar via oxygen plasma etching. After annealing in H$_2$/Ar gas at 350 ºC to remove resist residue [17], the device was mounted on a cryostat in an ultra-high vacuum (UHV) chamber. The widths $w$ for all devices are 0.9 um and the lengths $L$ between two Hall bar junctions ranged from 1.8 to 3.2 um. All measurements were taken by lock-in techniques at a low frequency of 3.7 Hz. A high input impedance (1 TΩ at dc) voltage preamplifier was used to eliminate artifacts in the nonlocal measurement [9].

Adatoms (Au, Ir) were evaporated on graphene devices via electron-beam evaporator in UHV. To vary the density of adatoms, the device was exposed to a controlled flux in sequential time intervals at sample temperature of 7 K. Figure 1(b) shows the gate-voltage dependence of the local conductivity $\sigma_{xx}(V_g) = [\rho_{xx}(V_g)]^{-1}$ for pristine and Au-decorated graphene with $L/w$ = 2.9. For pristine graphene, the gate voltage of minimum conductivity $V_{g, min}$ is located at 45 V and the mobility $\mu$ is only ~ 5000 cm$^2$/Vs, indicating the as-fabricated graphene device is strongly $p$-type doped and rather disordered. Upon cooling from 300 K to 7 K, the overall shape of $\sigma_{xx}(V_g)$ is the same with a slight increase in mobility. The adatom doping gradually shifts the $V_{g, min}$ to more negative gate voltage ($n$-type doping), lowers the mobility to ~ 1000 cm$^2$/Vs and decreases the minimum conductivity $\sigma_{min}$. Figure 1(c) shows the shift of $V_{g, min}$ as a function of Au adatom coverage. At low doping level, the shift in $V_{g, min}$ is roughly linear in Au coverage, with an estimated charge transfer of ~0.005 $e$ per Au adatom, while at higher coverages, the shift of $V_{g, min}$ is sublinear in Au coverage. As the coverage is on order 1 ML the Au is likely clustered [18, 19], with each cluster transferring on order one charge. Then the sublinearity reflects a growth in cluster size. The reduction of mobility with shift in $V_{g, min}$ [Fig. 1(c) inset] is quantitatively similar to that of the previous work using potassium adatoms on graphene [20], indicating the Au adatoms act as charged impurities.

We also studied the temperature dependence of $\rho_{xx}$ of graphene with Au and Ir adatoms to search for an energy gap induced by spin orbit coupling [15] or inversion symmetry breaking [8]. Figure 1(d) shows $\rho_{xx}(T)$ for graphene with Au adatom decoration. $\rho_{xx}(T)$ is very weakly insulating, rising slightly faster than logarithmically with decreasing temperature. $\rho_{xx}(T)$ is poorly described by a simple thermal activation model $\rho_{xx} \propto e^{E_g/k_B T}$; i.e. no activated behavior is seen (see Supplemental Material Fig. S1 [21]), and we conclude any gap is much smaller than the disorder scale set by electron-hole puddling, on order 50 meV.

Figure 2(a) presents the gate-voltage dependence of $R_{NL}$ for pristine and Au-decorated graphene device with $L/w = 2.9$. A peak in $R_{NL}(V_g)$ is evident for both pristine and Au-doped graphene and shifts toward negative $V_g$ after doping, which is consistent with the shift of $V_{g,min}$. However, the peak magnitude does not increase with increasing Au adatom density, although the local resistance $\rho_{xx}$ increases significantly. The width of the $R_{NL}(V_g)$ peak becomes broader after doping. $R_{NL}(V_g)$ for pristine graphene shows little temperature dependence, having almost the same amplitude in 300 K and 7 K [Fig. 2(a)] and is also much greater than the expected Ohmic contribution to the non-local resistance $R_{NL,Ohmic} = \rho_{xx} e^{-\pi L/w}$ (~1 Ω). We also verified that $V_{NL}$ is linear in $I_{NL}$, ruling out a thermoelectric source [5] of $R_{NL}$ (see Supplemental Material Fig. S2 [21]). Figure 2(b) shows the ratio of $R_{NL}/\rho_{xx}$ as a function of shifted gate voltage ($V_g - V_{g,min}$) for pristine and Au-decorated graphene. $R_{NL}/\rho_{xx}$ is not constant (as would be expected for the Ohmic contribution) but rather strongly peaked near $V_{g,min}$ as has been observed previously for neutral Hall currents in graphene [7, 8]. However, in this case $R_{NL}/\rho_{xx}$ exhibits a noticeable electron-hole asymmetry not previously observed. We find that the asymmetric peak can be well fit by Breit-Wigner-Fano (BWF) resonance function: $\frac{R_{NL}}{\rho_{xx}} = A_0 + A_1 \frac{[q+\tilde{V}_g]^2}{1+\tilde{V}_g^2}$, where $A_1$ is the amplitude of the resonance, $q$ is the asymmetry parameter, where we define a dimensionless gate voltage $\tilde{V}_g \equiv \frac{V_g - V_{g,min} - V_{g,res}}{\Gamma}$, where $\Gamma$ is the width.

Figure 3 shows the BWF fitting parameters for pristine and Au-decorated graphene at various $L/w$ ratios. (The gate-voltage dependence of $R_{NL}$ and $R_{NL}/\rho_{xx}$ for pristine and Au-decorated graphene with other $L/w$ ratios are shown in the Supplemental Material Fig. S3 [21]). Figures 3 (a-c) show the fitting parameters $A_1$, $q$ and $\Gamma$ as a function of mobility. The amplitude $A_1$ is roughly independent of mobility [Fig. 3(a)] while the asymmetry parameter $q$ is roughly proportional to mobility [Fig. 3(b)]. The width of the BWF peak $\Gamma$ decreases with increasing mobility, and the magnitude is roughly consistent with the dependence of the width of the minimum conductivity region on mobility as observed for charged impurity disorder [20, 22]. Figure 3(d) shows $A_1$ depends exponentially on length with similar magnitude and decay length for several $L/w$ electrode pairs on three pristine graphene devices. We conclude that there is no qualitative difference in behavior between pristine and Au-decorated graphene. For Ir-decorated graphene devices, we observed very similar features (See Supplemental Material Figs. S4 and S5 [21]): Ir adatoms also act as charged impurities and donate electrons to graphene (~0.01 $e$ per atom). The ratio of $R_{NL}/\rho_{xx}$ as a function of $V_g$ - $V_{g,\,min}$ is also well fit by BWF resonance function, with similar trend of the fitting parameters with disorder and aspect ratio.

For a neutral Hall effect it can be shown that [7, 8]:

$$\frac{R_{NL}}{\rho_{xx}} = \frac{1}{2}(w/\xi)e^{-L/\xi}\gamma_n^2 \qquad (1)$$

where $\gamma_n$ is the neutral Hall angle and $\xi$ is the scattering length. By fitting the data in Fig. 3(d) to Eq. (1), we find $\xi$ is consistently in the range of 250-400 nm. This $\xi$ value is consistent with the measured intervalley scattering length in exfoliated graphene flakes on $SiO_2$ [23] and is much shorter than the measured spin scattering lengths of 3 - 12 um for graphene [24] [note that the data in Fig. 3(d) correspond to as-fabricated devices with no adatoms that might modify the spin-orbit scattering]. We have also measured the parallel magnetic field dependence $B_{//}$ of $R_{NL}$ for as-fabricated graphene, as shown in Fig. 4. We find a small negative MR of less than 10%, consistent with a small weak localization effect due to parallel field penetrating the corrugations of graphene on $SiO_2$ [25]. If $\gamma_n$ represented a SHE then spin precession in parallel magnetic field (the Hanle effect [26]) would cause $R_{NL}(B_{//})$ to oscillate with field and reverse sign (see Supplemental Material Fig. S6 [21]). Assuming that $R_{NL}$ originates from SHE and the spin scattering length $\lambda_s = \xi = 300$ nm, we find the spin relaxation time $\tau_s = 1.2$ ps. The red curves in Fig. 4 show the corresponding expected $R_{NL}(B_{//})$ due to Hanle spin precession. We also show $R_{NL}(B_{//})$ for a hypothetical device with $\lambda_s = 3000$ nm and $\tau_s = 100$ ps (comparable to that measured by other groups [24]). No evidence of precession up to a field of 6 T is seen. The lack of precession indicates that the neutral current is not a spin current induced by SHE. Taken together, these features indicate that the observed neutral Hall effect is due to VHE, not SHE; $\gamma_n = \gamma_v$.

Comparing the BWF formula to Eq. (1), we find $\gamma_v^2 \propto \frac{[q + \tilde{V}_g]^2}{1 + \tilde{V}_g^2}$. This suggests two contributions to $\gamma_v$: One contribution has $|\gamma_{v,1}| \propto \frac{|\tilde{V}_g|}{\sqrt{1 + \tilde{V}_g^2}}$ and the other has $|\gamma_{v,2}| \propto \frac{q}{\sqrt{1 + \tilde{V}_g^2}} \propto \frac{\mu}{\sqrt{1 + \tilde{V}_g^2}}$. The fact that the contributions are added and then squared, indicates that they result from the same type of neutral current, which must be valley. The VHE requires inversion symmetry breaking. We hypothesize that disorder breaks inversion symmetry locally, leading to a VHE. While a microscopic theory of disorder induced VHE is lacking, we compare to the results for graphene with spin-orbit coupling. Sinitsyn *et al* [27] studied graphene with a spin-orbit gap, and found for correlated disorder a constant Hall angle independent of the disorder potential strength and of the concentration of scatters. Resonant scatters (skew scattering) in graphene with spin-orbit coupling tend to give Hall angles that are inversely proportional to the impurity concentration, large for Fermi energies near the resonance energy, and antisymmetric in Fermi energy about the resonance. This suggests that $\gamma_{v,1}$ (with magnitude roughly independent of $V_g$) results from correlated charge disorder, is an even function of $\tilde{V}_g$, and dominates in the highly charge-disordered sample, while $\gamma_{v,2}$ (with magnitude increasing with decreasing $V_g$) results from resonant scattering from mid-gap states induced by skew scatters, is an odd function of $\tilde{V}_g$, and is most important in relatively clean samples.

One important question is: What determines the overall sign of $\gamma_{v,1}$ and $\gamma_{v,2}$? The experimental setup probes only $\gamma_v^2$ since the injection and detection of the valley current are accomplished through the same mechanism, and each is proportional to $\gamma_v$. However opposite signs of $\gamma_v$ in the injector and detector would produce an overall negative $R_{NL}$, which is never observed in 16 injector/detector pairs on 4 samples. Disorder induced by atomically sharp defects, such as structural defects, chemisorbed species, and substitutional defects breaks the hexagonal symmetry of the honeycomb lattice and results in intervalley scattering [28, 29]. However, there is no reason to expect global favoring of A or B sublattice in the device, and one would expect a random mixture of inversion-symmetry-broken phases with cancelling contributions to $\gamma_v$. One possibility is that the disorder is highly correlated, due to interactions between disorder sites which tend to order the system [30-33]. Another possibility is that the sign of $\gamma_v$ is determined by another factor, such as out-of-plane symmetry breaking by an electric field. It is possible that point defects result from bonding to the $SiO_2$ substrate, breaking inversion symmetry out of plane. The charge disorder in as-fabricated devices results in *p*-type doping, likely due to negative charges in the substrate, while the adatoms on top of graphene result in *n*-type doping. In both cases the electric field is vertically into the substrate. We have indirect evidence that the disorder/electric field due to impurities may be important: We found a device with very small $V_{g, min}$ near 0 ($|V_{g, min}|< 0.5$

V) and mobility >10000 cm$^2$/Vs; the $\sigma_{xx}(V_g)$ is shown in Supplemental Material Fig. S7(a) [21]. In contrast to the previous devices with $V_{g, min} > 40$ V, this device exhibited no measurable gate dependent $R_{NL}$ (Supplemental Material Fig. S7(b) [21]) for a $L/w$ ratio of 3.2.

As shown in Fig. 2(a), $R_{NL}$ has near the same amplitude and shape at both 300 K and 7 K. The lack of temperature dependence reflects the energy scale of disorder induced VHE is very large. The energy difference between the A and B sublattices for a point defect such as a vacancy is locally on order the bandwidth, ~7.5 eV [34]. For a Coulomb impurity situated 3 Å above atom A, the energy difference between the A and B sublattices is ~0.5 eV. In both cases the energy scales are much greater than room temperature, which, if the disorder may be understood and controlled, provides an effective method for making room temperature valleytronic devices in graphene.

Lastly, we note that a very recent experimental study of $R_{NL}$ in hydrogenated graphene [11] found a similar decay length $\xi$ and lack of temperature and magnetic field dependence as seen here. This work casts doubt on the previous interpretation of $R_{NL}$ in hydrogenated graphene as due to SHE [6, 7], and opens the possibility of a disorder-induced VHE in that system as well.

References

[1] A. Rycerz, J. Tworzydlo, and C. W. J. Beenakker, Nat. Phys. **3**, 172 (2007).

[2] D. Xiao, W. Yao, and Q. Niu, Phys. Rev. Lett. **99**, 236809 (2007).

[3] D. A. Abanin, S. V. Morozov, L. A. Ponomarenko, R. V. Gorbachev, A. S. Mayorov, M. I. Katsnelson, K. Watanabe, T. Taniguchi, K. S. Novoselov, L. S. Levitov, and A. K. Geim, Science **332**, 328 (2011).

[4] D. A. Abanin, R. V. Gorbachev, K. S. Novoselov, A. K. Geim, and L. S. Levitov, Phys. Rev. Lett. **107**, 096601 (2011).

[5] J. Renard, M. Studer, and J. A. Folk, Phys. Rev. Lett. **112**, 116601 (2014).

[6] A. H. Castro Neto and F. Guinea, Phys. Rev. Lett. **103**, 026804 (2009).

[7] J. Balakrishnan, G. Kok Wai Koon, M. Jaiswal, A. H. Castro Neto, and B. Özyilmaz, Nat. Phys. **9**, 284 (2013).

[8] R. V. Gorbachev, J. C. W. Song, G. L. Yu, A. V. Kretinin, F. Withers, Y. Cao, A. Mishchenko, I. V. Grigorieva, K. S. Novoselov, L. S. Levitov, and A. K. Geim, Science **346**, 448 (2014).

[9] M. Sui, G. Chen, L. Ma, W. Shan, D. Tian, K. Watanabe, T. Taniguchi, X. Jin, W. Yao, D. Xiao, and Y. Zhang, arXiv:1501.04685.

[10] Y. Shimazaki, M. Yamamoto, I. V. Borzenets, K. Watanabe, T. Taniguchi, and S. Tarucha, arXiv:1501.04776.

[11] A. A. Kaverzin and B. J. van Wees, Phys. Rev. B **91**, 165412 (2015).

[12] D. Huertas-Hernando, F. Guinea, and A. Brataas, Phys. Rev. B **74**, 155426 (2006).

[13] H. Min, J. E. Hill, N. A. Sinitsyn, B. R. Sahu, L. Kleinman, and A. H. MacDonald, Phys. Rev. B **74**, 165310 (2006).

[14] Y. Yao, F. Ye, X.-L. Qi, S.-C. Zhang, and Z. Fang, Phys. Rev. B **75**, 041401 (2007).

[15] J. Hu, J. Alicea, R. Wu, and M. Franz, Phys. Rev. Lett. **109**, 266801 (2012).

[16] J. Balakrishnan, G. K. W. Koon, A. Avsar, Y. Ho, J. H. Lee, M. Jaiswal, S.-J. Baeck, J.-H. Ahn, A. Ferreira, M. A. Cazalilla, A. H. C. Neto, and B. Özyilmaz, Nat Commun **5**, 4748 (2014).

[17] M. Ishigami, J. H. Chen, W. G. Cullen, M. S. Fuhrer, and E. D. Williams, Nano Lett. **7**, 1643 (2007).

[18] K. M. McCreary, K. Pi, A. G. Swartz, W. Han, W. Bao, C. N. Lau, F. Guinea, M. I. Katsnelson, and R. K. Kawakami, Phys. Rev. B **81**, 115453 (2010).

[19] Y. Wu, W. Jiang, Y. Ren, W. Cai, W. H. Lee, H. Li, R. D. Piner, C. W. Pope, Y. Hao, H. Ji, J. Kang, and R. S. Ruoff, Small **8**, 3129 (2012).

[20] J. H. Chen, C. Jang, S. Adam, M. S. Fuhrer, E. D. Williams, and M. Ishigami, Nat. Phys. **4**, 377 (2008).

[21] See Supplemental Material for additional data.

[22] S. Adam, E. H. Hwang, V. M. Galitski, and S. Das Sarma, Proc. Natl. Acad. Sci. U.S.A. **104**, 18392 (2007).

[23] F. V. Tikhonenko, D. W. Horsell, R. V. Gorbachev, and A. K. Savchenko, Phys. Rev. Lett. **100**, 056802 (2008).

[24] W. Han, R. K. Kawakami, M. Gmitra, and J. Fabian, Nat Nano **9**, 794 (2014).

[25] M. B. Lundeberg and J. A. Folk, Phys. Rev. Lett. **105**, 146804 (2010).

[26] M. Johnson and R. H. Silsbee, Phys. Rev. B **37**, 5312 (1988).

[27] N. Sinitsyn, J. Hill, H. Min, J. Sinova, and A. MacDonald, Phys. Rev. Lett. **97**, 106804 (2006).

[28] V. V. Cheianov and V. I. Fal'ko, Phys. Rev. Lett. **97**, 226801 (2006).

[29] E. McCann, K. Kechedzhi, V. I. Fal'ko, H. Suzuura, T. Ando, and B. L. Altshuler, Phys. Rev. Lett. **97**, 146805 (2006).

[30] V. V. Cheianov, O. Syljuåsen, B. L. Altshuler, and V. Fal'ko, Phys. Rev. B **80**, 233409 (2009).

[31] V. V. Cheianov, V. I. Fal'ko, O. Syljuåsen, and B. L. Altshuler, Solid State Commun. **149**, 1499 (2009).

[32] V. V. Cheianov, O. Syljuåsen, B. L. Altshuler, and V. I. Fal'ko, EPL **89**, 56003 (2010).

[33] D. A. Abanin, A. V. Shytov, and L. S. Levitov, Phys. Rev. Lett. **105**, 086802 (2010).

[34] F. Banhart, J. Kotakoski, and A. V. Krasheninnikov, ACS Nano **5**, 26 (2010).

Figure Captions :

Figure 1. (a) AFM image of our Hall-bar geometry graphene device. The width $w$ and length $L$ of the Hall bar, and the configuration of current $I_{NL}$ and voltage probes $V_{NL}$ for measurement of the non-local resistance $R_{NL} = V_{NL}/I_{NL}$ are indicated. (b) The conductivity $\sigma_{xx}$ versus gate voltage $V_g$ curves for graphene with $L/w = 2.9$ for as-fabricated graphene and at three different Au adatom concentrations. Here 1 ML = 1.4 x $10^{15}$ cm$^{-2}$ for Au(111). (c) The shift of gate voltage of minimum conductivity -$\Delta V_{g,min}$ as a function of Au coverage. The inset shows the -$\Delta V_{g,min}$ vs. inverse mobility; here all -$\Delta V_{g,min}$ values are offset by 10 V to account for the initial disorder in the sample. Lines correspond to the theory developed in Ref. 20. (d) Temperature dependence of $\rho_{xx}$ at $V_g = V_{g,min}$ point for Au-decorated graphene.

Figure 2. (a) Non-local resistance $R_{NL}$ versus $V_g$ curves for as-fabricated and Au-decorated graphene with $L/w = 2.9$. The dashed and solid black curves show the measurement of as-fabricated graphene at 300 K and 7 K, respectively. (b) The ratio of $\frac{R_{NL}}{\rho_{xx}}$ as a function of shifted gate voltage $V_g$ -$V_{g,min}$ for as-fabricated and Au-decorated graphene with $L/w = 2.9$. The solid lines are a fit to a Breit-Wigner-Fano function as described in text.

Figure 3. (a) - (c) The parameters of the Breit-Wigner-Fano function $A_1$ (a), $q$ (b) and $\Gamma$ (c) as a function of mobility at various $L/w$ ratios. Parameters are described in text. (d) The fitting parameter $A_1$ for three different as-fabricated graphene devices as a function of length.

Figure 4. $R_{NL}$ for as fabricated graphene device with $L = 1.4$ um and $w = 0.9$ um at $V_g - V_{g,min} = 5$ V as a function of parallel magnetic field $B_{\parallel}$. The squares and black line are the measured data, the dotted black line is the calculated Ohmic contribution and the red lines are calculated Hanle precession for $\lambda_s = 300$ nm, $\tau_s = 1.2$ ps (solid red line) and $\lambda_s = 3000$ nm, $\tau_s = 100$ ps (dashed red line).

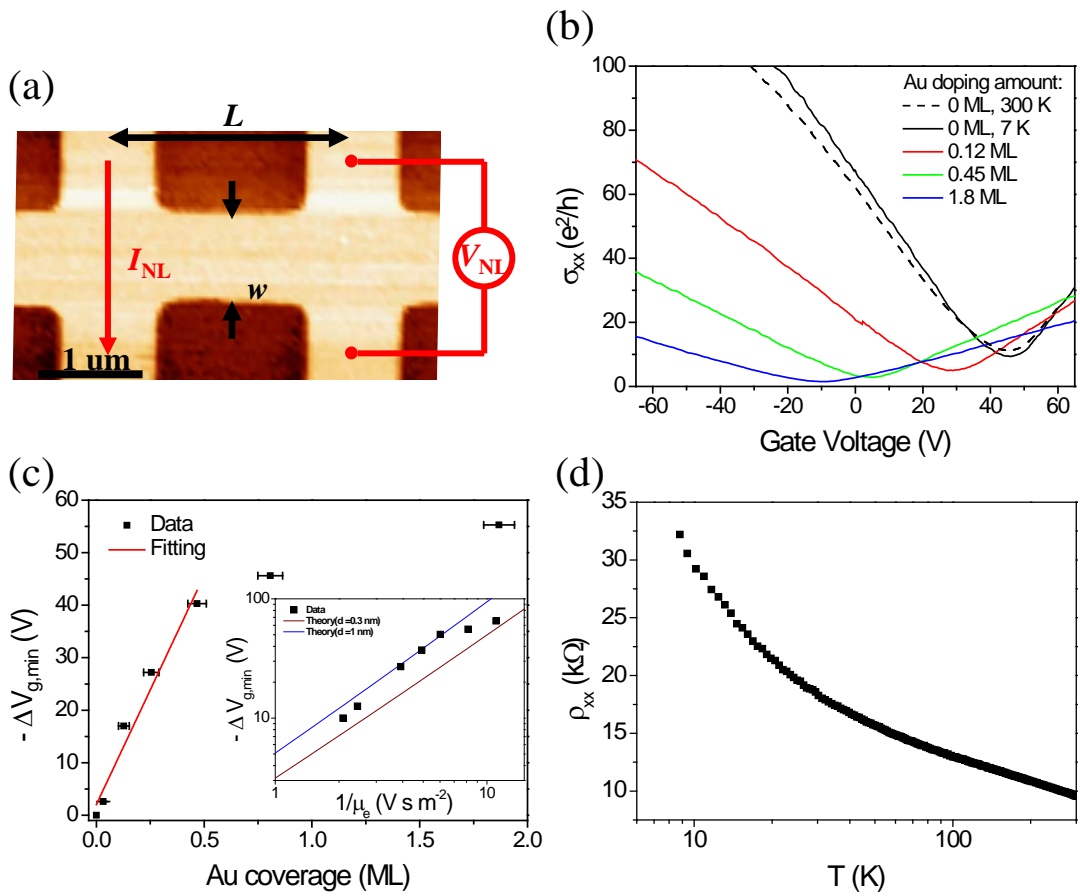

Fig. 1, Y.L. Wang *et al*

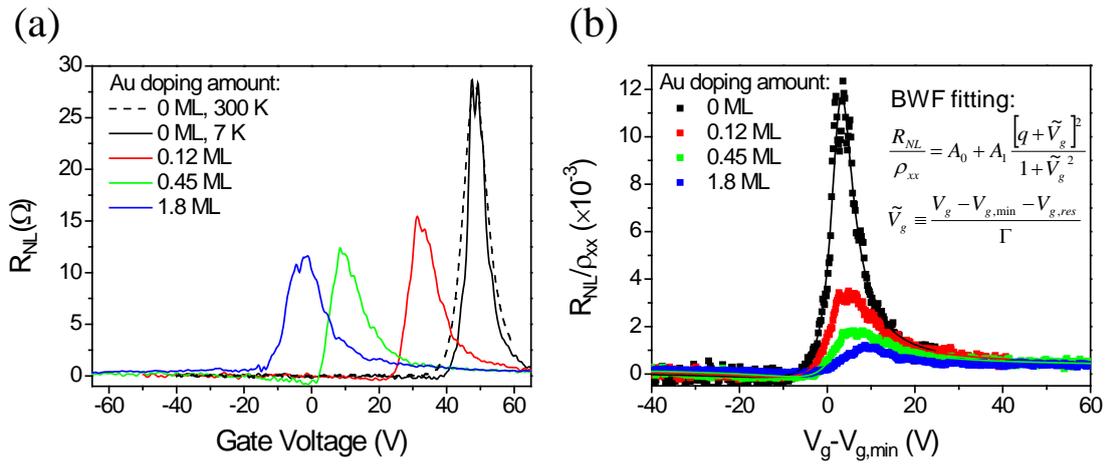

Fig. 2, Y.L. Wang *et al*

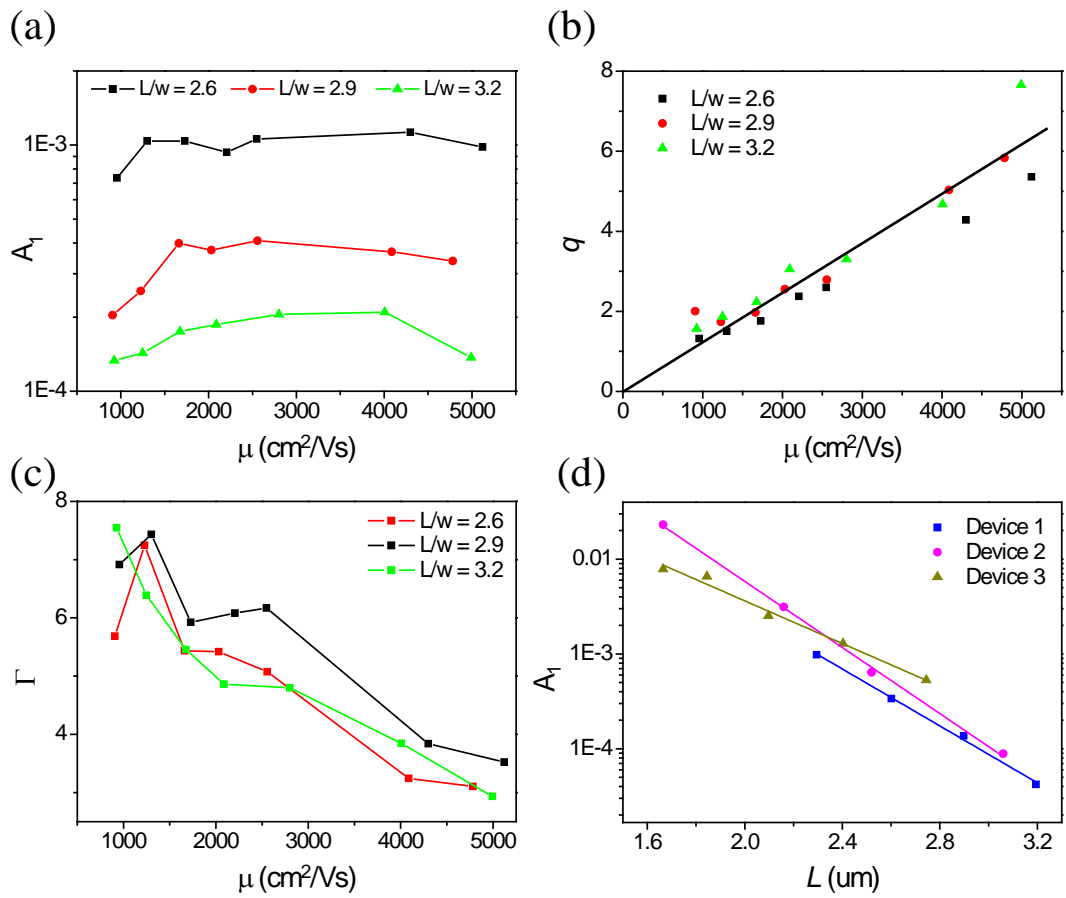

Fig. 3, Y.L. Wang *et al*

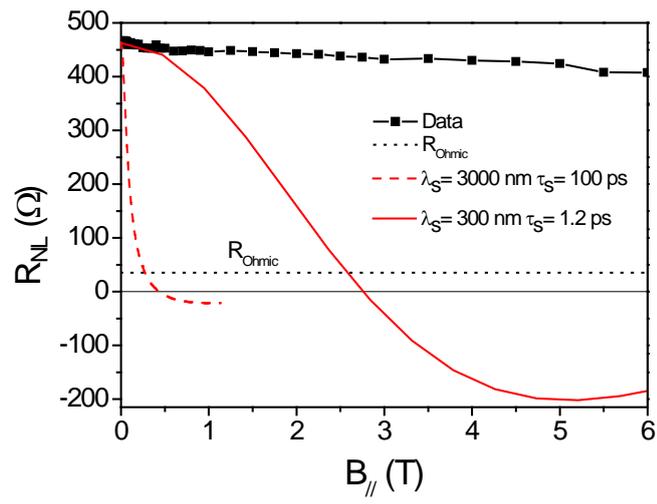

Fig. 4, Y.L. Wang *et al*

Supplemental Material for "Neutral-current Hall effects in disordered graphene"

Yilin Wang,[1,2,4] Xinghan Cai,[2] Janice Reutt-Robey[1,4] and Michael S. Fuhrer[1,2,3*]

[1]*Materials Research Science and Engineering Center, Department of Physics, University of Maryland, College Park, MD 20742, USA*

[2]*Center for Nanophysics and Advanced Materials, University of Maryland, College Park, MD 20742, USA*

[3]*School of Physics, Monash University, Victoria 3800, Australia*

[4]*Department of Chemistry and Biochemistry, University of Maryland, College Park, MD 20742, USA*

* michael.fuhrer@monash.edu

**Lack of activated behavior of $\rho_{xx}$**

We plot $\rho_{xx}$ in logarithmic scale as a function of inverse temperature and fit the data with equation $\rho_{xx} \propto \exp(E_g / k_b T)$ in the low temperature region, in order to get the thermally activated gap. The obtained fitting gap is extremely small, $E_g < 1$ meV, which is unphysically small in that it is smaller than the measurement temperature $k_B T$ and much smaller than the disorder energy scale of order 50 meV. We speculate that the failure to observe the predicted large energy gap for isolated Ir on graphene is because Ir form clusters on graphene [1], which is different from the single adatom model used in the theory [2].

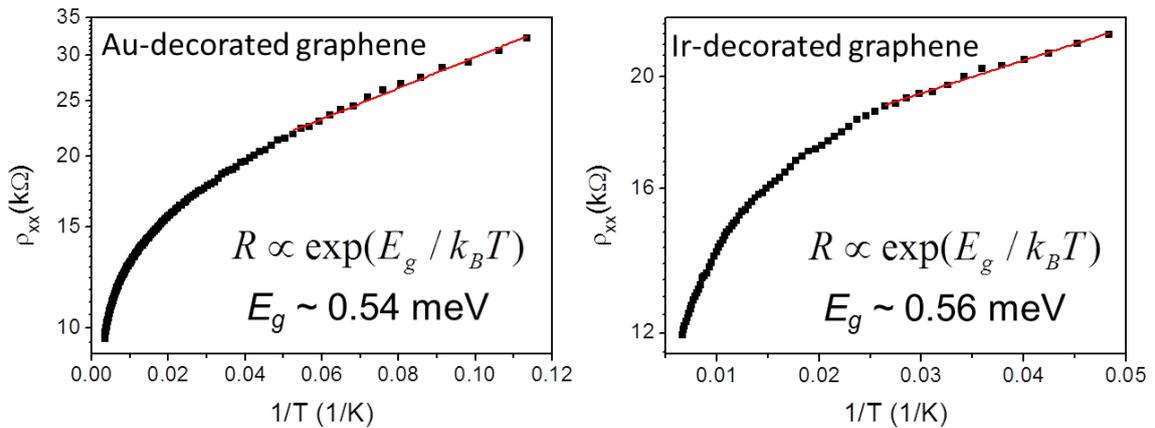

FIG. S1 Temperature dependence of $\rho_{xx}$ at $V_g = V_{g,\min}$ point for Au and Ir-decorated graphene. The red line is a fit to the thermal activation model as described in the text.

**Linearity of the non-local signal**

We measure $V_{NL}$ at different applied currents $I_{NL}$, as shown in Fig. S2(a). Figure S2(b) shows the dependence of $R_{NL} = V_{NL}/I_{NL}$ on gate voltage. We find that the magnitude of $R_{NL}$ is independent of the current $I_{NL}$ indicating that $V_{NL}$ is linear in $I_{NL}$. This excludes thermoelectric effects due to non-uniform Joule heating as the origin of $R_{NL}$.

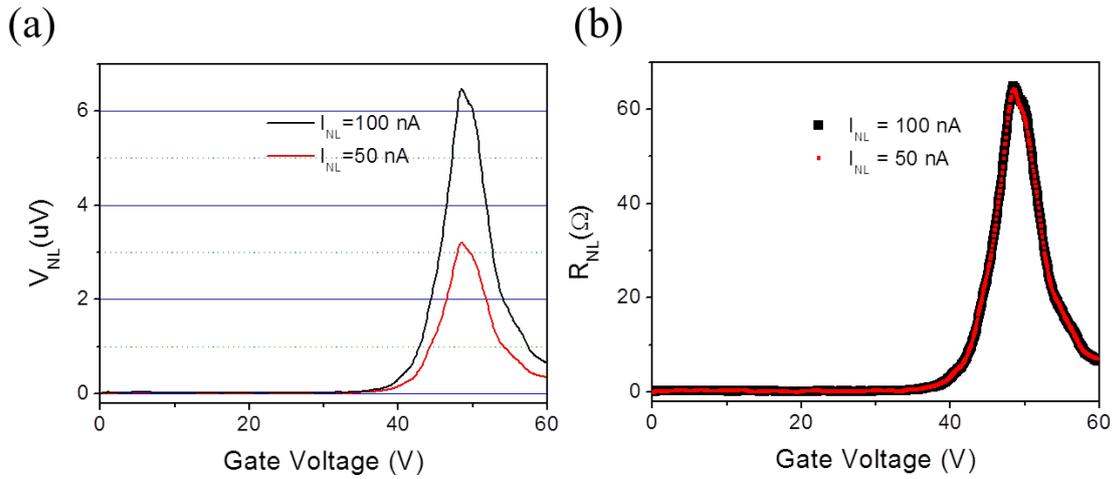

FIG. S2 The $V_{NL}(V_g)$ dependence (a) and $R_{NL}(V_g)$ dependence (b) for device with $L/w$=2.6 at different current magnitudes as indicated in legend.

**The non-local resistance for pristine and Au-decorated graphene at various *L/w* ratios**

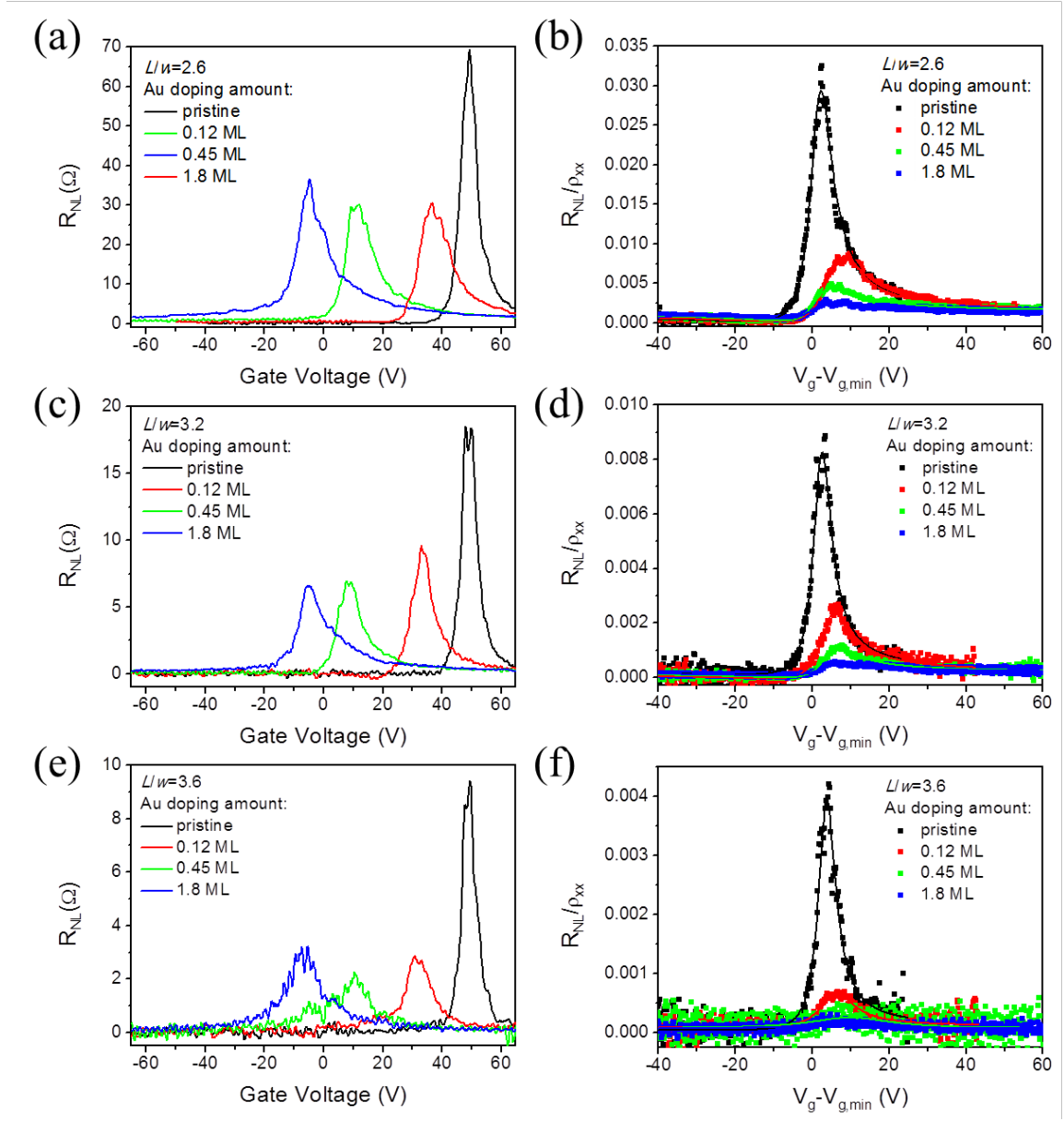

FIG. S3. The non-local resistance $R_{NL}$ as a function of gate voltage $V_g$ and the ratio of $\frac{R_{NL}}{\rho_{xx}}$ as a function of shifted gate voltage $V_g - V_{g,min}$ for pristine and Au-decorated graphene at the *L/w* ratio of 2.6 (a) and (b), 3.2 (c) and (d) and 3.6 (e) and (f). Solid lines are fits to the Breit-Wigner-Fano function as described in the main text.

**Ir adatoms act as charged impurities on graphene**

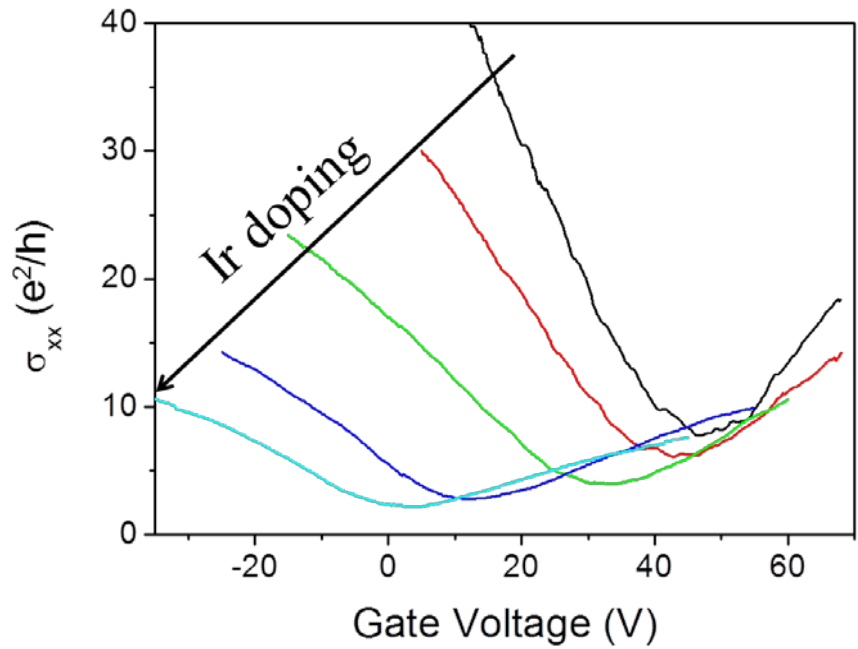

FIG. S4. The conductivity $\sigma_{xx}$ versus gate voltage $V_g$ curves for the pristine and Ir-decorated graphene. Similar to the case of Au adatom deposition (discussed in main text), Ir adatom deposition donates electron to graphene, shifting $V_{g,min}$ to the negative voltage direction, and the mobility and minimum conductivity decrease with increasing Ir doping.

**The non-local resistance for Ir-decorated graphene at various *L/w* ratios**

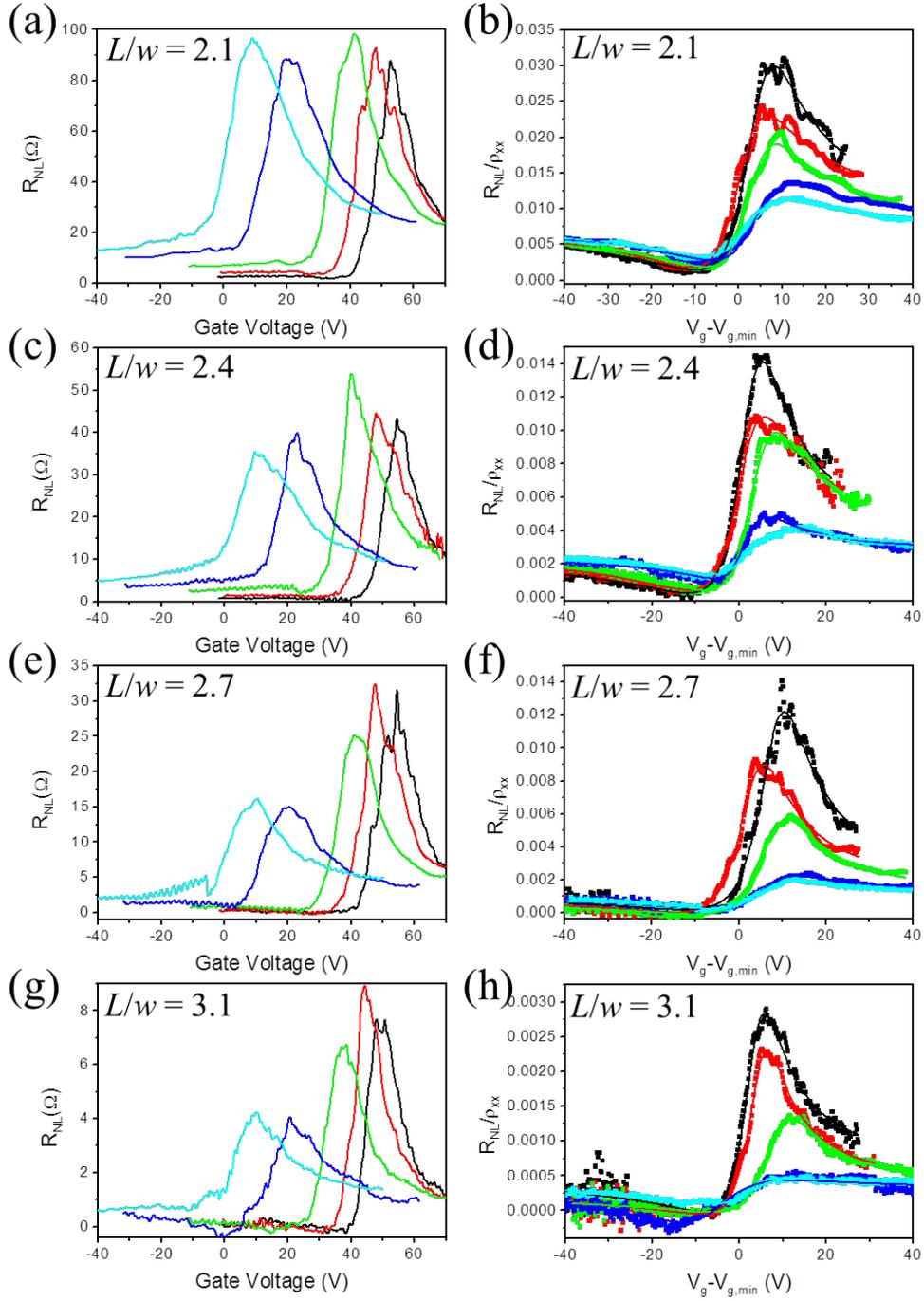

FIG. S5. The non-local resistance $R_{NL}$ as a function of gate voltage $V_g$ and the ratio of $\frac{R_{NL}}{\rho_{xx}}$ as a function of shifted gate voltage $V_g - V_{g,min}$ for pristine and Ir-decorated graphene at the *L/w* ratio of 2.1 (a) and (b), 2.4 (c) and (d) and 2.7 (e) and (f) and 3.1 (g) and (h). Solid lines are fits to the Breit-Wigner-Fano function as described in the main text.

**Precession analysis in parallel magnetic field**

For the spin Hall effect, $R_{NL}$ should undergo precession in parallel magnetic field. The Hanle spin precession can be expressed in the formula [3]:

$$R_{NL} \propto \frac{1}{f(\omega_B \tau_s)} \left[ \sqrt{1+f(\omega_B \tau_s)} \cos\left[\frac{l\omega_B \tau_s}{\sqrt{1+f(\omega_B \tau_s)}}\right] - \frac{\omega_B \tau_s}{\sqrt{1+f(\omega_B \tau_s)}} \sin\left[\frac{l\omega_B \tau_s}{\sqrt{1+f(\omega_B \tau_s)}}\right] \right] e^{-l\sqrt{1+f(\omega_B \tau_s)}},$$

where $f(\omega_B \tau_s) = \sqrt{1+(\omega_B \tau_s)^2}$ and $l = L/\sqrt{2\lambda_s^2}$, $\omega_B = \gamma B$ is the Larmor frequency, $\gamma$ is the gyromagnetic ratio, B is the magnetic field, $\tau_s$ is the spin relaxation time and $\lambda_s$ is the scattering length. If we assume the observed $R_{NL}$ in our devices originates from spin Hall effect, $\lambda_s$ is in the range of 250 nm - 400 nm. Figure S5 shows the expected Hanle precession for $\lambda_s$ =300 nm (black curve); $R_{NL}$ oscillates and becomes negative at specific magnetic field. A more pronounced oscillation is seen if we assume that $\lambda_s$ =3000 nm, similar to other observations in pristine graphene on $SiO_2$ [4].

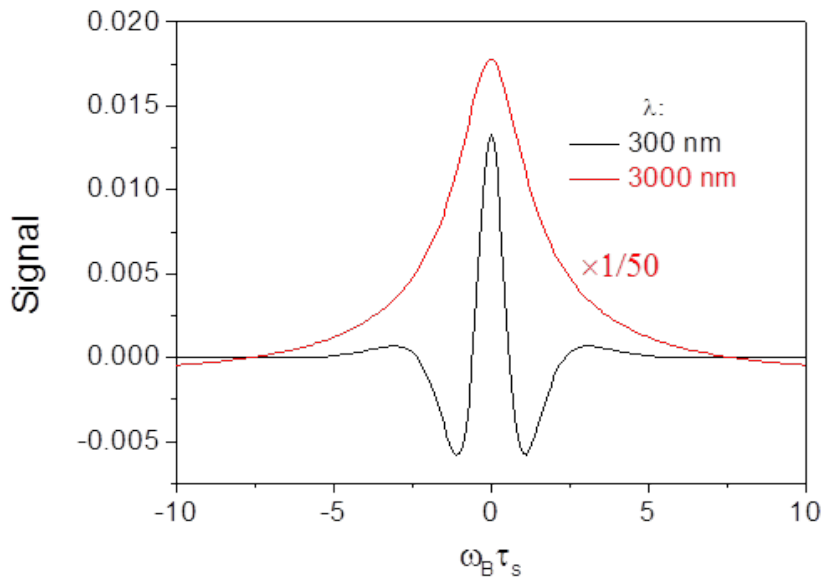

FIG. S6. The calculated Hanle precession curves when $\lambda_s$ is 300 nm and 3000 nm.

**No measurable gate dependent $R_{NL}$ in low-doped graphene device**

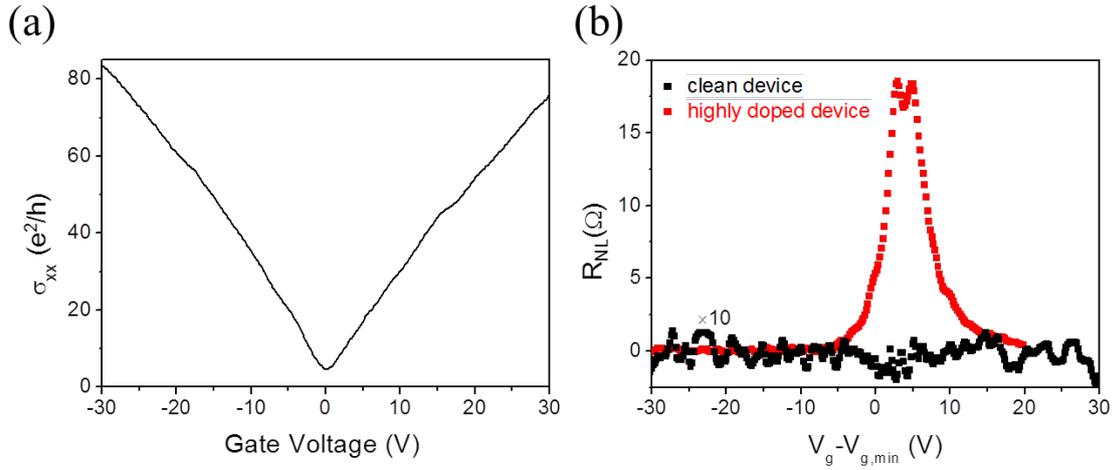

FIG. S7. (a) The $\sigma_{xx}$ versus $V_g$ curve for an as-fabricated graphene device with $L/w$=3.2, and unusually low doping: $|V_{g,\,min}|$< 0.5 V and $\mu$ >10000 cm$^2$/Vs. (b) The $R_{NL}$ versus $V_g - V_{g,min}$ curves for the as-fabricated clean graphene and highly doped device with the same $L/w$=3.2. No measurable gate-dependent $R_{NL}$ is seen in the low-doped device.

**Supplemental References:**

1   B. A. Barker, A. J. Bradley, M. M. Ugeda, S. Coh, A. Zettl, M. F. Crommie, M. L. Cohen, and S. G. Louie, Bull. Am. Phys. Soc. **60** (2015).
2   J. Hu, J. Alicea, R. Wu, and M. Franz, Phys. Rev. Lett. **109**, 266801 (2012).
3   M. Johnson and R. H. Silsbee, Phys. Rev. B **37**, 5312 (1988).
4   W. Han, R. K. Kawakami, M. Gmitra, and J. Fabian, Nat Nano **9**, 794 (2014).